\documentclass[prl, onecolumn, 12pt, superscriptaddress]{revtex4}
\usepackage{graphicx}
\usepackage{colortbl}
\usepackage{amsmath,amssymb}
\usepackage{rotating}

\usepackage[light,first,bottomafter]{draftcopy}

\newcommand{\PR}{{\em Phys. Rev. }}
\newcommand{\PRL}{{\em Phys. Rev. Lett. }}
\newcommand{\PL}{{\em Phys. Lett. }}
\newcommand{\NP}{{\em Nucl. Phys. }}
\newcommand{\NIM}{{\em Nuclear Instruments and Methods}}
\newcommand{\etal}{{\em et al}}

\def \pr{\mbox{$r_p$}}

\begin{document}
\title{To the measurement of the radius via an electron scattering}
\author{B.~Wojtsekhowski} \thanks{bogdanw@jlab.org}
\affiliation{\mbox{Thomas Jefferson National Accelerator Facility, Newport News, VA 23606}}

\begin{abstract}
We propose an experiment for an accurate measurement of the proton radius.
A key feature of our proposal is an iron-free magnetic spectrometer.  
Projected systematics uncertainties will allow a 1\% level accuracy for the \pr{} value. 
\end{abstract}

\maketitle

\section{Introduction}
The absolute value of the proton electromagnetic radius is a subject of significant interest~\cite{poh10, ber10, joh15, dou15}.  
Current knowledge of the proton radius is summarized in the PDG report~\cite{pdg10}.  
The recommended value of \pr{} is $0.877 \pm 0.007$~fm. 
Recent results of the $\mu$-hydrogen spectroscopical experiment~\cite{poh10} 
suggest that electron-proton scattering could have unaccounted systematics.
A number of potential interpretations have been discussed, see e.g.~\cite{dis11} and references within.
Measuring the proton radius is a difficult task due to the very small size
of the effect that the proton size introduces in the form factor at the momentum transfer
when the notion of rms charge distribution is applicable with the required accuracy.

\section{The concept}
The measurement of the proton form factor at low momentum transfer is a
textbook method of electromagnetic radius determination.
The form factor, $F_p$, is defined by the equation: \\
\begin{center}
$F_p^2 \,=\, \frac{N_{counts}}{N_p \cdot N_e} \times \frac{4 E_i^2}{r_e^2}
\frac{sin^4(\theta_e/2)}{\Delta \Omega}$,
\end{center}
where $N_{counts}$ is the number of elastic scattering events (after
correction for radiative effects), $N_p$ is the number of protons per cm$^2$
in the target area visible by a detector in the beam path, 
$N_e$ is the number of beam electrons passing through the target during the data taking,
$E_i$ is the incident electron beam energy,  $r_e$ is the electron
classical radius, $\theta_e$ is the scattering angle, and $\Delta \Omega$ is 
the detector solid angle.

The form factor is related to the rms of the proton charge distribution.  
At small momentum transfer in the first Born approximation, $F_p \,=\, 1 - b^2/6$, where
$b^2 = Q^2 \cdot r_p^2$ is a product of the square of the momentum transfer, $Q^2 \,=\, 4
E_i \cdot E_f \cdot \sin^2 (\theta_e/2)$ and the squared radius, \pr$^2$. 
Model dependence of the \pr{} determination becomes acceptably low
only  at $b \ll 1$.
Even at a very small $b$, a number of known corrections need to be applied.
Among them are corrections due to the two-photon effects.

Numerous electron scattering experiments have confirmed 
that $b$ of $\sim 0.1$ is sufficiently low for a 1\% level determination of the radius.
For example, the Carbon-12 radius measurement was performed by using 
the form factor measurement at $b$ of $\sim 0.1$~\cite{car80}.
It is interesting to note that the muon X-ray technique result for the
$r_{_{Carbon-12}}$ radius~\cite{sch82} is in perfect agreement with the
one from the electron scattering.
At $b \,=\, 0.1$, the form factor is equal to 0.9983.  
The uncertainty $\delta r_p/r_p \approx \delta b/b$ is 300 times
larger than $\delta F_p$.
For 1\% accuracy in the radius, the form factor needs to be measured to
a precision of $3 \cdot 10^{-5}$.  The cross section measurement should
have an accuracy better than $6 \cdot 10^{-5}$.

All experimental parameters need to be determined with very high
accuracy for the proton radius measurement.  They are:
\begin{itemize}
\item Electron beam energy
\item Electron scattering angle
\item Electron beam intensity
\item Proton target thickness
\item Solid angle of the electron spectrometer
\item Efficiency of the detector in the spectrometer
\end{itemize}

Some of these parameters could be measured rather well:
The photon back scattering from the multi-MeV electron provides an
approach for the precision beam energy measurement.
The compensated calorimeter method could be used in the precise
measurement of the beam power in the beam intensity range of a few $\mu$A.
However, currently achieved accuracies for other
parameters are not sufficient for the \pr{} experiment.
Therefore, instead of an absolute determination of the form factor, we are
considering a measurement of the form factor variation with 
the momentum transfer.
The measurement of the form factor difference at $b = 0.05$ and $b =
0.15$ could be realized with a high accuracy.
In such an approach the stability of the apparatus becomes very important.
At the same time, much larger uncertainties of the absolute values e.g.
of the target areal density and the detector solid angle would be acceptable.

The measurements at several values of $b$ could be achieved by the adjustment of the
spectrometer position and its momentum.
However, each of these operations would result in a big uncertainty in 
the setup parameters.  
For example, relocation of the spectrometer will change
the spectrometer solid angle and effective thickness of the target.

We propose here a different way to change $b$: a variation of the beam 
energy with a fixed location of the spectrometer.  
In such a case, only the momentum settings need to be adjusted.
In the traditional magnetic spectrometers used in the electron scattering experiments, 
a change in the spectrometer momentum may lead to a significant change in the solid angle.
Calibration of the solid angle with the required precision presents a
difficult problem especially for the case of the extended hydrogen target.
We propose to construct a custom magnetic spectrometer for the momentum
range up to 75~MeV without the use of ferromagnetic material.

\section{The proposed experimental setup}

\begin{figure}[htb]
\centerline{\includegraphics[width=0.8\textwidth]{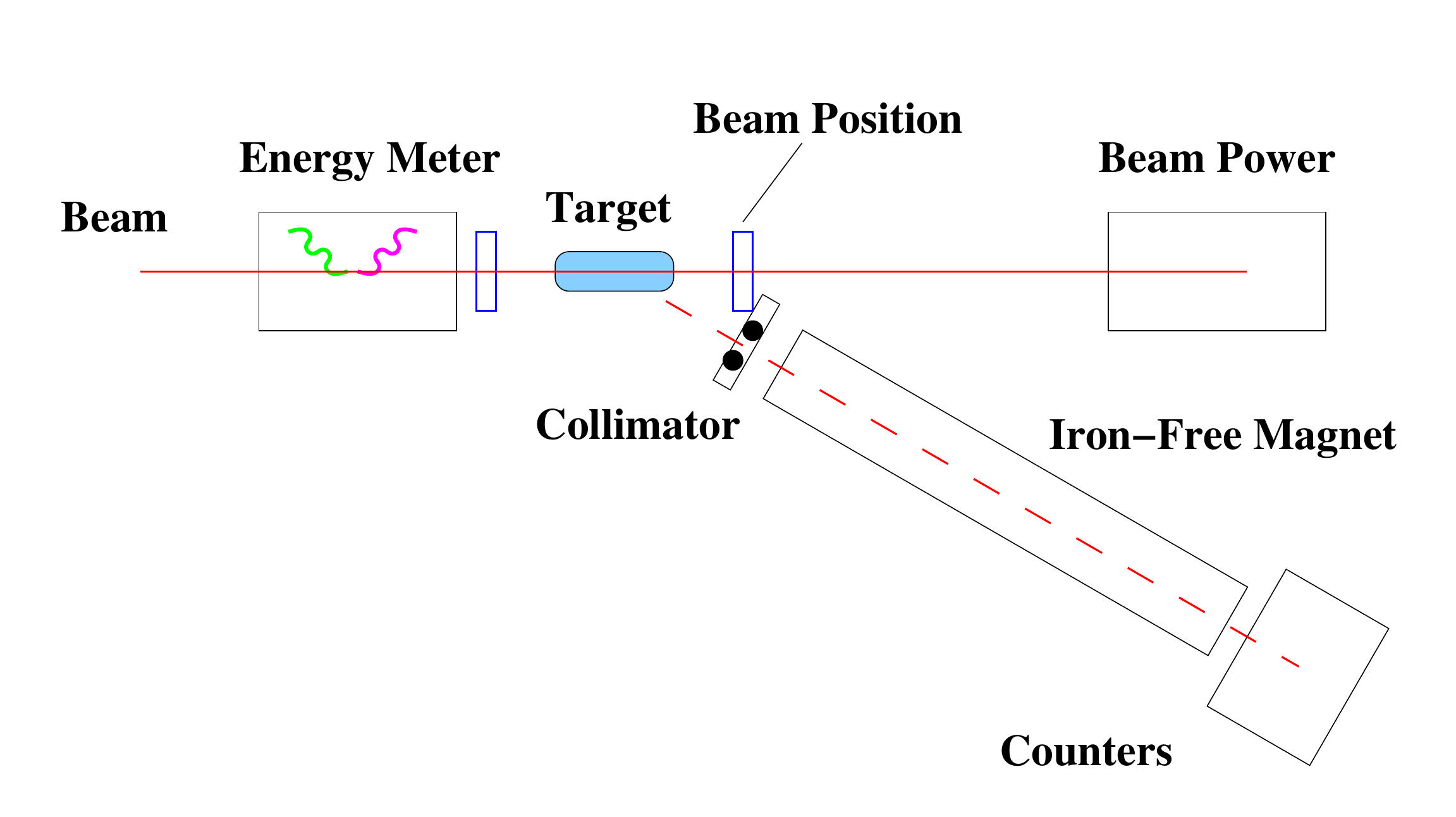}}
\caption[]{\label{fig:setup} The layout of the proposed experiment.}
\end{figure}

The layout of the proposed experiment is presented in Fig.~\ref{fig:setup}.
We would like to use an electron beam of 25-75 MeV and a scattering from a gas
hydrogen target in a thin-walled container.
A spectrometer will be located at a fixed angle of 30$^\circ$ relative to the beam direction.
The spectrometer will have a solid angle of 1~msr and transverse target acceptance of 0.5~cm.  

At these beam energies, scattering from the proton could only be
elastic.  Counting rates will be between 1 and 10~kHz for  a 1~$\mu$A beam 
with the hydrogen target at room temperature and with a pressure of 7.5~atm.

\subsection{Electron beam}

There are several accelerators which could provide a low energy beam
with the required parameters.  
They are the CEBAF injector, the JLab FEL, the MAMI A, and DALINAC.  
The beam energy measurements will be accomplished with a laser
back-scattering system.  
The energy spectrum of the backscattered X-rays has an end-point in a keV range
which allows a precision calibration using radioactive sources.  
The total beam power is about 25-75 Watts.
It should be measured with a calorimeter with active stabilization
of its temperature.  The power of the heating element will allow
the measurement of the beam intensity.
  
\subsection{Hydrogen target}

A target pressure of 7.5~atm will allow the use of a thin-walled cell 
with carbon foil windows.  
With a cell diameter of 2.5~cm, the wall thickness will be of
5~$\mu$m.  The temperature will be stabilized by the usual means of a
heating element.  The pressure in the cell will be recorded.   
The electron energy loss in such a target will have a spread of 1~keV, 
which should be taken into account in the spectrometer momentum
setting.  Total energy deposition by the beam in the target will be about 20~mW.

\subsection{Detector}

The detector in the focal plane of the magnetic spectrometer
will include three or four planes of thin scintillator counters
followed by a thick counter of 30~cm. 
The data will be analyzed on-line via FADC electronics for
the trigger and recorded for off-line analysis.
 
\subsection{Magnetic spectrometer}

For the low energy electrons of 4~MeV, the iron-free spectrometers were constructed 
with a solid angle of up to 35~msr, see e.g. Ref.~\cite{sie64}.  
With superconducting coils the iron-free toroidal spectrometer
was constructed for the GeV momentum range~\cite{ome89}.
However, room temperature coils allow some design advantages,
e.g. direct control of the coil geometry during an experiment.
In the proposed spectrometer a magnet will provide a magnetic field of 3~kG. 
A bend radius of 80~cm will allow for a compact device.  
A solid angle of 1~msr and momentum acceptance of 5\% should be 
achievable for a 0.5~cm target size (normal to the spectrometer mid-plane).
The in-plane angular acceptance should be several times smaller than
the out-of-plane one to facilitate control of the average scattering angle.
NMR probes will be used to check the proportionality between the field 
in the spectrometer and the current.  
The accuracy of such a measurement could be as good as $10^{-5}$ or even better.
The fringe field of the spectrometer will deflect the electron beam
incident on the target.  
However, the momentum of the incident electrons and 
the spectrometer momentum setting are almost equal to one
another.  Therefore, the beam path variation should be small and correctable.

\section{Discussion}

The study of the electron-proton scattering at low-$Q^2$ has been very
active recently, see e.g.~\cite{ber10} and references within.  
Connection of the form factors to the light-cone densities
at large values of an impact parameter through the dispersion theory 
has been elaborated in Ref.~\cite{mil11}.
Accurate measurement of the proton form factor ratio, $G_E^p/G_M^p$,  down to 
a small momentum transfer of 0.22~GeV$^2$ is reported in Ref.~\cite{ron11}.
A concept of $G_E^p/G_M^p$ measurement at even smaller $Q^2$ by means of 
a proton beam was introduced in Ref.~\cite{ron09}.  
The nucleon form factor data analysis via the dispersion relations~\cite{bel07} 
results in a value of \pr =0.844~fm, which is in good agreement with the
recent muon-hydrogen Lamb shift result~\cite{poh10}, see also~\cite{dou15}.
There is significant interest in resolving concern about the validity of 
the low $Q^2$ electron-proton elastic scattering method of \pr{} determination.

In summary, we propose a concept of an experiment to measure a proton form factor 
momentum transfer dependence at very small values of the momentum transfer
which should be capable of resolving the current proton size ``crisis''.
The experimental apparatus under consideration could also be used for 
the measurements of the radius of other nuclei.

\end{document}